\documentclass[11pt,twoside]{article}

\usepackage{asp2014}

\aspSuppressVolSlug
\resetcounters

\begin{document}

\title{PyCPL: The ESO Common Pipeline Library in Python v1.0}

\author{Mrunmayi S. Deshpande,$^1$ Nuria P. F. Lorente,$^1$ Anthony Horton,$^1$ Brent~Miszalski,$^1$ Ralf Palsa,$^2$ Lars Lundin,$^2$ Anthony Heng,$^1$ Aidan Farrell$^1$}
\affil{$^1$Australian Astronomical Optics, Macquarie University, Sydney, Australia; \email{Mrunmayi.Deshpande@mq.edu.au}}
\affil{$^2$European Southern Observatory, Garching, Germany}

\paperauthor{Mrunmayi S. Deshpande}{Mrunmayi.Deshpande@mq.edu.au}{ORCID_Or_Blank}{Macquaire University}{Australian Astronomical Optics}{Sydney}{NSW}{2109}{Australia}
\paperauthor{Nuria P. F. Lorenete}{Author2Email@email.edu}{ORCID_Or_Blank}{Macquaire University}{Australian Astronomical Optics}{Sydney}{NSW}{2109}{Australia}
\paperauthor{Anthony Horton}{Anthony.Hortan.edu.au}{ORCID_Or_Blank}{Macquaire University}{Australian Astronomical Optics}{Sydney}{NSW}{2109}{Australia}
\paperauthor{Brent Miszalski}{Brent.Miszalski@mq.edu.au}{ORCID_Or_Blank}{Macquaire University}{Australian Astronomical Optics}{Sydney}{NSW}{2109}{Australia}
\paperauthor{Ralf Palsa}{rpalsa@eso.org}{ORCID_Or_Blank}{European Southern Observatory}{Author3 Department}{Garching}{Garching bei München}{85748}{Germany}
\paperauthor{Lars Lundin}{llundin@eso.org}{ORCID_Or_Blank}{European Southern Observatory}{Author3 Department}{Garching}{Garching bei München}{85748}{Germany}
\paperauthor{Anthony Heng}{Anthony.Heng@mq.edu.au}{ORCID_Or_Blank}{Macquaire University}{Australian Astronomical Optics}{Sydney}{NSW}{2109}{Australia}
\paperauthor{Aidan Farrell}{Aiden.Farrell@mq.esu.au}{ORCID_Or_Blank}{Macquaire University}{Australian Astronomical Optics}{Sydney}{NSW}{2109}{Australia}



\begin{abstract}
PyCPL provides full access to ESO’s Common Pipeline Library (CPL) for astronomical data reduction within a Python environment. Not only does it offer a Python interface to the robust CPL library, but it also lets users and developers fully utilise the rest of the scientific Python ecosystem. We have written a C++ layer to CPL and with pybind11 (a third-party library) created a Pythonic API to CPL. Since CPL has been around for so long, it has been thoroughly tested and understood. In 2003 it was developed in C due to its efficiency and speed of execution. With the community however moving away from C/C++ programming and embracing Python for data processing tasks, there is a need to provide access to the CPL utilities within a Python environment. With the latest version being released users can now install PyCPL to run existing CPL recipes (written in C) and access the results from Python. It also provides the ability to create new recipes in Python using the functionality provided by CPL.
\end{abstract}



\section{Introduction}

\subsection{The Common Pipeline Library}
The Common Pipeline Library (CPL) is a collection of ISO-C libraries that offer a strong, fast, and extensive software toolkit for creating astronomical data-reduction tasks, or recipes. These can either be executed manually by the user or initiated by an automated framework (called "pipelines").

The Common Pipeline Library \citep{2004ASPC..314..764J} was created to standardise the building process of VLT instrument pipelines, reducing the duration of their development and simplifying their maintenance. The CPL code has been engineered to limit external third party dependencies and make the library portable and adaptable, and it can be used outside the VLT context for similar applications.

CPL provides a host of functionalities, presented in a clear, generic and uniform manner. It is used at ESO for a quick look at data processing at the observatory. CPL is also used for data product creation from the ESO archive facility, and monitoring of the status of VLT instruments \citep{2011vlt1612}.
Since its initial release in 2003, CPL has been an essential tool for developing data reduction pipelines for ESO VLT instruments. It will continue to be used as the foundation for pipelines involving ELT instruments \citep{2011cplManual}.
CPL also forms the basis for ESO's High-level Data Reduction Library (HDRL) which implements more complex or specialised data processing algorithms.

\section{PyCPL}
As the science and engineering community are moving away from the C/C++ languages and adopting Python for data processing tasks, we have created PyCPL, a library that enables using CPL in a more Pythonic way. PyCPL offers Python language bindings for the CPL tool kit API, including the CPL plugin interface \citep{2011cplManual}. With PyCPL one can run existing CPL recipes and access the results from Python. The user can also create new recipes in Python by using the functionalities in CPL together with other Python libraries (e.g.~Astropy).
Through PyCPL one may develop Python recipes that comply with ESO standards and execute the recipes of currently available ESO instrument pipeline packages from Python scripts or the interpreter prompt. 
A command line program, PyEsoRex, can be used analogously with the C-based EsoRex and execute pipeline recipes that are already developed using the CPL C API, as well as those that are implemented using the PyCPL Python API.
This possibility of accessing CPL through Python was initially investigated for the MUSE instrument on ESO VLT using the PYTHON-CPL module, which made it possible to run previously produced CPL recipes \citep{2012ASPC..461..853S}.

\subsection{Bindings}
Python bindings to CPL are achieved by means of an intermediary C++ layer with the use of pybind11\footnote{
\href{https://github.com/pybind/pybind11}{https://github.com/pybind/pybind11}} , a lightweight header-only library that exposes C++ types in Python and vice versa \citep{pybind11}. The bindings provide the necessary user-facing Python interface to our C++ layer (see Figure \ref{fig:PyCPL architecture}). Python provides an easy platform to build and manage recipes, especially when it comes to changing recipe parameters for simulations. 

With the advent of Python bindings, we can improve the accessibility of CPL. This ranges from the most basic length() and dump() functions to some more complex functions including array based access to \texttt{cpl.core.Images}. In some special cases we can even make use of Numpy masked arrays that allow us to record invalid elements efficiently.

\subsection{Three Layer Model}
\begin{figure}
    \centering
    \includegraphics[width = 0.9\linewidth]{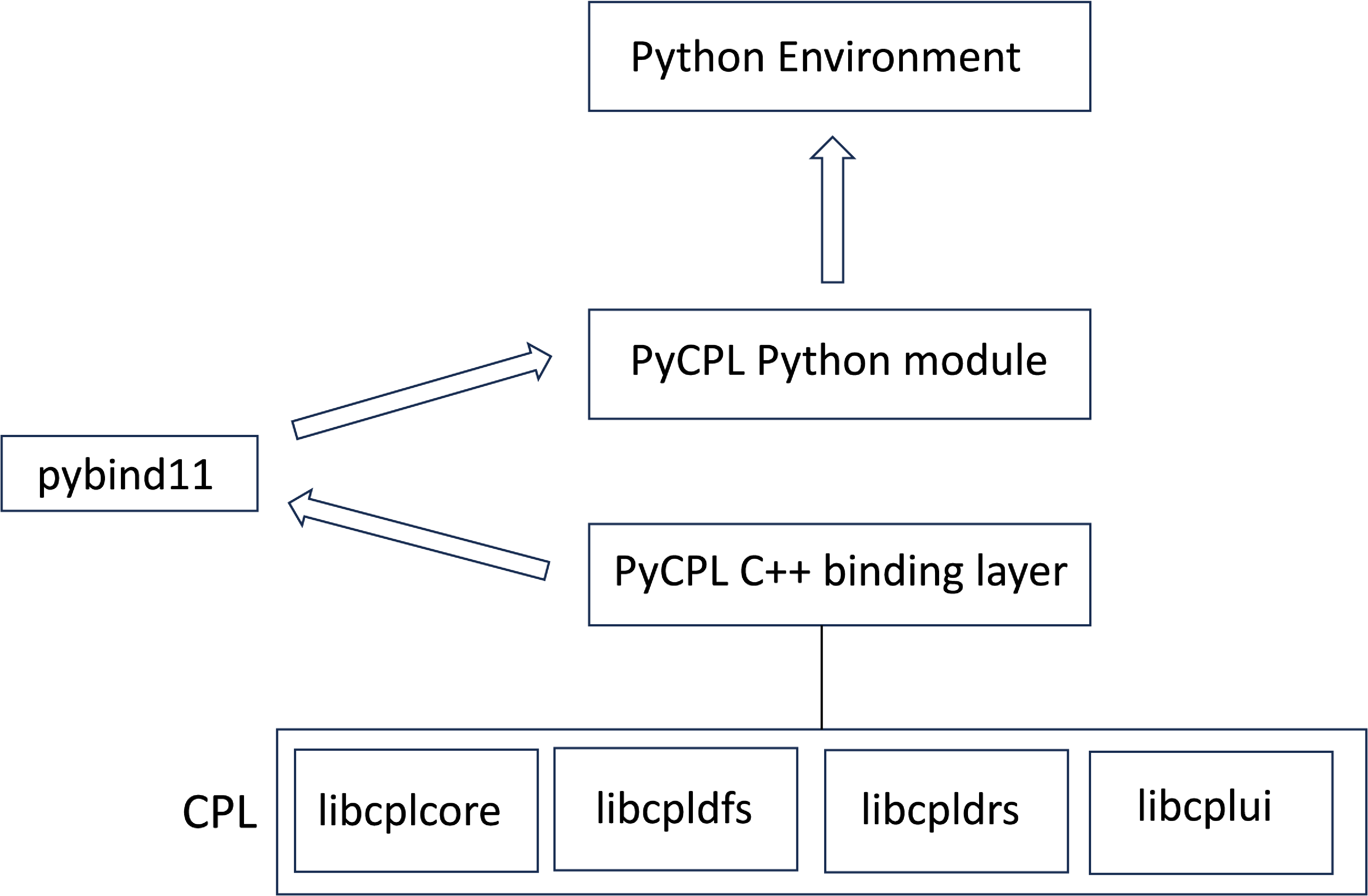}
    \caption{PyCPL architecture layout with three layers namely CPL modules, bindings layers and python environment.}
    \label{fig:PyCPL architecture}
\end{figure}

Figure \ref{fig:PyCPL architecture} shows the three layer model used to create PyCPL. 
Even with the addition of pybind11, straight translation from CPL to PyCPL does not occur automatically. We have created a three-layer model that wraps the existing CPL ISO-C code. 
Creating an object-oriented C++ layer is the first step. All
the CPL library's features are included in the PyCPL libraries, divided into four smaller modules \citep{Horton2023}:
\begin{enumerate}
  \item
  The CPL library itself, written in ISO-C in a pseudo object-oriented style (structs with sets of associated functions). 
  The CPL library contains C object-like constructs.
  \begin{itemize}
    \item \texttt{cpl.core}: Classes containing the core modules of image, mask, table, matrix, vector, metadata, statistics,
arithmetic, I/O and conversion operations for these classes, function fitting and exceptions.
    \item \texttt{cpl.dfs}:  This module contains data flow system related functions for integrating data products into the ESO Data Flow System.
    \item \texttt{cpl.drs}: Classes Containing functions for data reduction software, higher level tools for WCS, plate solving, geometric transformations, wavelength
calibration.
    \item \texttt{cpl.ui}: This module consists of classes representing recipes, parameters, data frames, and lists of data frames.
\end{itemize}
  \item The second layer consists of the C++ binding layer. This translates C \texttt{structs} and functions to C++ classes and methods as an object-oriented wrapper for the CPL C library. The C++ layer’s goal is to turn CPL \texttt{structs} and the functions that go along with them like tables, properties, and images into C++ classes, for a better integration with the pybind11 Python binding generator.
  \item The Python Interface which is the third layer, provides access to the CPL library in Python. The C++ classes and methods are mapped to natural Python classes and methods using the Python bindings.
\end{enumerate}

\subsection{Testing}
The ESO CPL library is robust, well-tested and has been extensively used since 2003. With PyCPL we have adopted and expanded the CPL tests, creating unit tests for each PyCPL function. The PyCPL source repository therefore includes a comprehensive set of unit, validation and regression tests.

\section{PyCPL version 1.0}
PyCPL version 1.0 was released on 6th November 2023.
It comes with numerous API changes as compared to the previous versions. It can be obtained from the ESO \mbox{Common} Pipeline Library web page~\footnote{
\href{https://www.eso.org/sci/software/pycpl/}{https://www.eso.org/sci/software/pycpl/}}.

\section{PyCPL Applications}

PyCPL provides Pythonic interfaces and the advantage of utilising a well understood, robust C library in Python.
Our first application of PyCPL, developed separately at the AAO, is an extensible data reduction framework that uses PyCPL and PyEsoRex to manage parameters and run a user-specified list of recipes. The regular framework provided by PyCPL substantially reduces the amount of code required to keep track of recipe input and output products, thus minimising maintenance and likelihood of errors. The framework currently includes MUSE data reduction workflows, and we plan to include more pipelines in the near future. 

We encourage our community to consider using PyCPL for data reduction and software pipeline development, and use it to prototype recipes and algorithms.

\bibliography{P923.bib}  

\end{document}